%% file: 2_comuns_v3.tex
\documentclass[aps,prb,reprint,twocolumn,superscriptaddress,floatfix,longbibliography]{revtex4-2}
\usepackage{graphicx}
\usepackage{float}
\usepackage{physics}
\usepackage{cancel}
\usepackage{overpic}
\usepackage{enumerate}
\usepackage{dsfont}
\usepackage{bbm}
\usepackage{mathrsfs}

\usepackage[colorlinks,citecolor=blue,linkcolor=blue,urlcolor=blue]{hyperref}
\usepackage{textcomp}
\usepackage{amsmath}
\usepackage{amssymb}
\usepackage{soul}
\usepackage[normalem]{ulem}
\usepackage{comment}

\usepackage[english]{babel}
\usepackage{bm}
\usepackage{orcidlink}

\graphicspath{{./}{./figs/}}

\include{shortcuts}

\begin{document}

%\title{Potts Models Universality Class in Driven Dissipative Bosonic Systems}

\title{Quantum and Classical Potts Criticality in Driven-Dissipative Bosonic Lattices}
\author{Jacopo Tosca~\orcidlink{0009-0002-5165-7937}}
\affiliation{Universit\'{e} Paris Cit\'e, CNRS, Mat\'{e}riaux et Ph\'{e}nom\`{e}nes Quantiques, 75013 Paris, France}

\author{Zejian Li~\orcidlink{0000-0002-5652-7034}}
\affiliation{Universit\'{e} Paris Cit\'e, CNRS, Mat\'{e}riaux et Ph\'{e}nom\`{e}nes Quantiques, 75013 Paris, France}

\author{Cristiano Ciuti \orcidlink{0000-0002-1134-7013}}
\affiliation{Universit\'{e} Paris Cit\'e, CNRS, Mat\'{e}riaux et Ph\'{e}nom\`{e}nes Quantiques, 75013 Paris, France}

\begin{abstract}

The emergence of equilibrium universality from intrinsically nonequilibrium dynamics is a fundamental open problem. Bose-Hubbard lattices realized in photonic and circuit-QED platforms provide a versatile setting to engineer nonlinear interactions, dissipation, and multiphoton processes. Here we investigate a Bose-Hubbard lattice subject to three-photon parametric driving, whose nonequilibrium steady state spontaneously breaks a $\mathbb Z_3$ symmetry and realizes the criticality of the three-state Potts model, a three-state generalization of the Ising model. Using a variational phase-space approach with systematically controllable accuracy based on a Variational Multi-Gaussian ansatz, we perform finite-size scaling analyses in one and two spatial dimensions. We find that, in two-dimensional lattices with single-photon losses, the nonequilibrium steady-state transition belongs to the universality class of the 2D classical three-state Potts model. In contrast, in one-dimensional lattices with three-photon losses, the transition is governed by the one-dimensional quantum three-state Potts universality class. These results establish driven-dissipative bosonic lattices as a platform for emergent Potts criticality and identify multiphoton dissipation as a mechanism that promotes nonequilibrium critical behavior from classical to quantum universality classes.

\end{abstract}

\maketitle

The emergence of critical phenomena in the nonequilibrium steady state of open quantum systems has attracted considerable attention in recent years~\cite{siebererUniversalityDrivenOpen2025,Breuer2002, Fazio2025}, especially in view of the possible realization of model systems via advanced quantum platforms such as circuit QED~\cite{houckOnchipQuantumSimulation2012a,carusottoPhotonicMaterialsCircuit2020,girvin2019schrodinger,blaisQuantumInformationProcessing2020a,PhysRevX.4.031043,Fitzpatrick_2017,pierangeli2019large,Beaulieu2025}, ultracold atoms~\cite{Muniz2020}, and photonic systems~\cite{Lebreuilly2017,Rota2019,Vicentini2019,tosca2025-EmergentEquilibrium,tosca2026isingselectormachinekerr,liDissipativePhaseTransition2022}, which can encode classical and quantum optimization problems in their nonequilibrium steady states~\cite{tosca2025-EmergentEquilibrium,tosca2026isingselectormachinekerr,liDissipationinducedAntiferromagneticlikeFrustration2021}.
 
Driven-dissipative photonic lattices have emerged as a versatile platform for exploring collective phenomena in open quantum many-body systems, raising the question of whether their nonequilibrium phase transitions can display universal behavior analogous to that of equilibrium statistical mechanics. A paradigmatic example is the quadratically-driven Bose-Hubbard model: as the two-photon driving is increased, the system spontaneously breaks a $\mathbb{Z}_2$ symmetry associated with photon-number parity~\cite{leghtas2015confining,wangSchrodingerCatLiving2016}, and this nonequilibrium transition was shown to exhibit emergent equilibrium criticality belonging to the quantum~\cite{Rota2019} or classical \cite{Verstraelen2020} Ising universality class depending on the presence or absence of two-photon losses. It is then intriguing to ask whether different universality classes can emerge in driven-dissipative systems with higher discrete symmetries. Beyond Ising, the archetypal such model is the $q$-state Potts model~\cite{Wu1982}, the canonical generalization of the Ising model to $q>2$ states. Its criticality is markedly rich: in two dimensions (2D) the transition is continuous only up to $q=4$ and turns first order beyond~\cite{Baxter1973}, while the continuous three-state case has non-Ising critical exponents and is governed by a $\mathbb{Z}_3$ parafermionic conformal field theory~\cite{Fateev1985}---the same parafermionic structure underlying proposals for topological quantum computation~\cite{Alicea2016}. Even if markedly present across statistical physics and combinatorial optimization, from adsorbed monolayers~\cite{Bretz1977} to percolation and the graph-coloring problem encoded in its antiferromagnetic ground states~\cite{Wu1982}, the Potts model has had no counterpart among the emergent universality classes of open quantum matter, which have remained confined to the $\mathbb{Z}_2$ Ising paradigm.
 
A promising route is offered by multiphoton-driven Kerr resonators. At the single-mode level, $n$-photon driven Kerr systems host a rich landscape of first- and second-order dissipative phase transitions~\cite{Gosner2020, Minganti2023,Kruglikov2025,brunoQuantumTheoryThreephoton2026}; the fate of the corresponding higher $\mathbb{Z}_n$ symmetries on extended lattices, however, has remained largely unexplored, and the case $n\ge3$ unaddressed out of the semiclassical regime \cite{Minganti2023}. The three-photon-driven case is especially compelling: it realizes a $\mathbb{Z}_3$ symmetry and has already been demonstrated experimentally in superconducting resonators, where a tri-squeezed state with triangular symmetry under $2\pi/3$-phase rotations was observed above threshold~\cite{Svensson2017,Chang2020}.
 
Characterizing symmetry breaking beyond the $\mathbb{Z}_2$ case is considerably more challenging. Three-photon driving and losses generate strongly correlated many-body states whose Wigner function develops pronounced negativities together with highly non-Gaussian features. Moreover, the presence of \emph{three}-photon pump makes the problem intractable for stochastic phase-space methods such as truncated-Wigner~\cite{Drummond1993,Polkovnikov_2010} and positive-$P$~\cite{Deuar2019}. Indeed, the presence of third-order phase-space derivatives, at the level of the Lindbladian, breaks completely the Fokker-Planck structure used for the mapping onto stochastic differential equations. 

In this Letter, we address these challenges with the Variational Multi-Gaussian (VMG) approach~\cite{Tosca2025VMG}, which represents the many-body Wigner function as a superposition of complex-centered Gaussian components with systematically controllable accuracy. Combining it with finite-size scaling analyses in 1D and 2D, we uncover emergent three-state Potts criticality in driven-dissipative bosonic lattices with a broken $\mathbb{Z}_3$ symmetry. We show that the 2D cubically-driven Bose-Hubbard model with single-photon losses undergoes a continuous transition in the universality class of the classical 2D three-state Potts model~\cite{Wu1982}, whereas a 1D chain in the presence of both single- and three-photon losses is governed by the 1D quantum three-state Potts universality class~\cite{rappDynamicalCorrelationsQuantum2006,rappAsymptoticScatteringDuality2013}.

\begin{figure*}[t]
\centering
\includegraphics[width=0.33\textwidth]{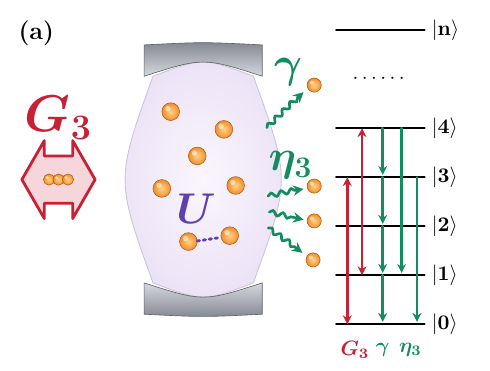}\hfill
\includegraphics[width=0.34\textwidth]{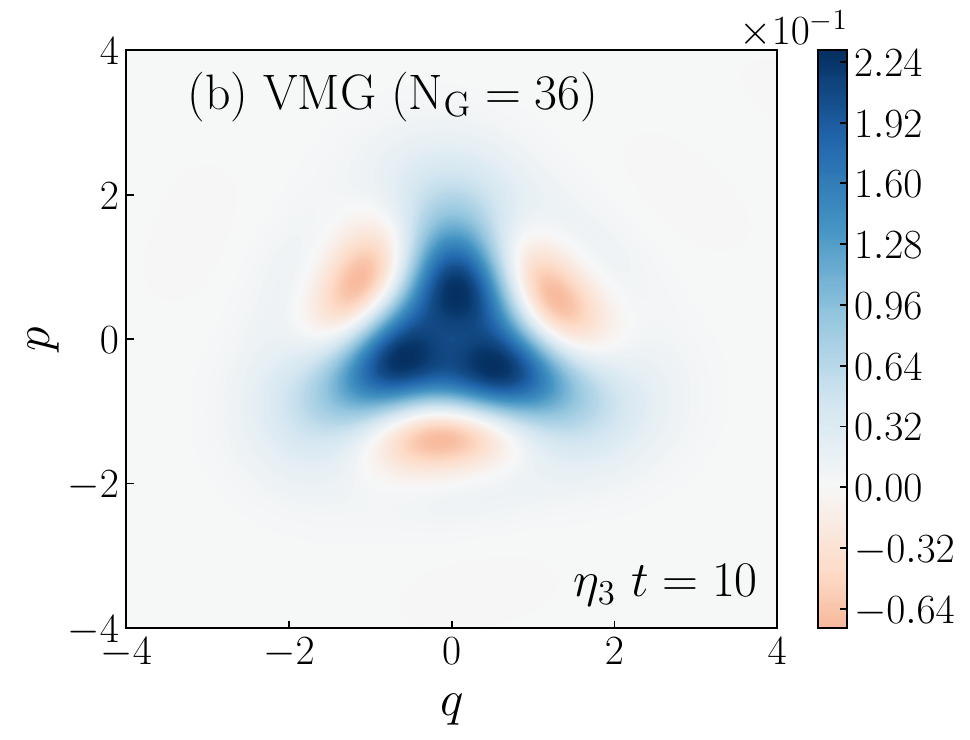}\hfill
\includegraphics[width=0.325\textwidth]{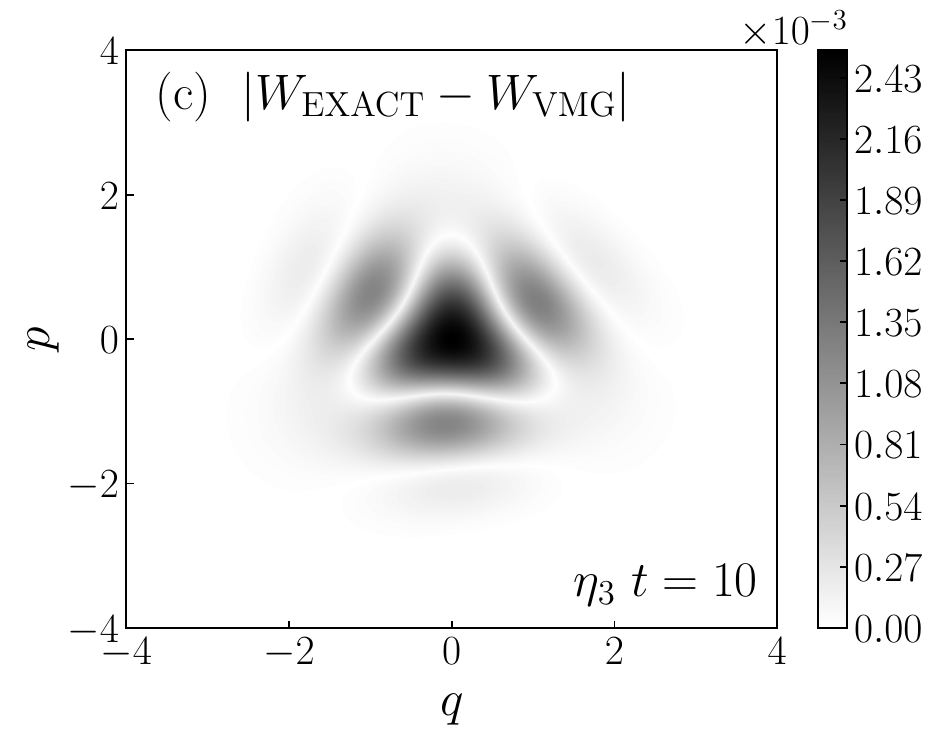}
\caption{
(a) Sketch of a photonic Kerr resonator with nonlinearity $U$, subject to three-photon parametric driving with amplitude $G_3$, single-photon losses at rate $\gamma$, and three-photon losses at rate $\eta_3$. The right panel illustrates the action of these processes on the Fock-state ladder.
(b) Snapshot of the Wigner function of the single-mode cubically driven dissipative Kerr parametric oscillator at time $t = 10/\eta_3$, obtained with the Variational Multi-Gaussian (VMG) ansatz using $12$ Gaussian triples (for a total of $N_G = 36$). System parameters: $G_3/\eta_3 = 0.75 + 0.75 i$, $U/\eta_3 = 1$ and $\Delta = \gamma = 0$.
(c) Absolute value of the difference between the exact and variational Wigner functions, demonstrating the high accuracy of the VMG ansatz in the presence of appreciable three-photon losses.
}
\label{fig:VMG_benchmark}
\end{figure*}
\textit{Model and Potts structure.}\,---\,A lattice of coupled quantum nonlinear optical resonators subject to 
three-photon parametric driving is described by the cubically-driven 
Bose-Hubbard model. In the frame rotating at one third of the pump 
frequency ($\hbar = 1$), the Hamiltonian reads
\begin{equation}\label{eq:Hamiltonian}
\begin{split}
\hat{H} &=\sum_j\!\left(-\Delta \hat{a}_j^{\dagger} \hat{a}_j
+\frac{U}{2} \hat{a}_j^{\dagger 2} \hat{a}_j^2
+ G_3 \hat{a}_j^{\dagger 3}+G_3^* \hat{a}_j^3\right)
\\
&-\sum_{\langle j, j^{\prime}\rangle} \frac{J}{z}
\left(\hat{a}_j^{\dagger} \hat{a}_{j^{\prime}}+\mathrm{h.c.}\right),
\end{split}
\end{equation}
where $\hat{a}_j$ is the photon annihilation operator on site~$j$ (so that $[\hat a_i , \hat a^\dagger_j] =\delta_{i,j}$), 
$U$~the Kerr nonlinearity and $G_3$~the three-photon driving amplitude; the parameter 
$\Delta = \omega_p/3 - \omega_c$ represents the detuning between the driving ($\omega_p$) and cavity frequency ($\omega_c$), while
$J$ is the nearest-neighbor hopping normalized by the coordination 
number~$z$. The dynamics of the system, when coupled to a Markovian bath, is governed by the Lindblad 
master equation~\cite{Breuer2002, Fazio2025}
\begin{equation}\label{eq:Lindblad}
\partial_t \hat{\rho} = \mathcal{L}\hat{\rho} 
= -i[\hat{H}, \hat{\rho}] + \sum_\mu \mathcal{D}[\hat{\Gamma}_\mu]\hat{\rho}\,,
\end{equation}
where $\mathcal D$ is the dissipation superoperator accounting for the  interaction channels with the external environment: 
\begin{equation}
\mathcal{D}[\hat{\Gamma}]\hat{\rho} = 
\hat{\Gamma}\hat{\rho}\hat{\Gamma}^\dagger 
- \frac{1}{2}\{\hat{\Gamma}^\dagger\hat{\Gamma},\hat{\rho}\}.
\end{equation}
In the following, we study the system in the presence of single- and three-photon losses with jump operators 
$\hat{\Gamma}_{1,j} = \sqrt{\gamma}\,\hat{a}_j$ and 
$\hat{\Gamma}_{3,j} = \sqrt{\eta_3}\,\hat{a}_j^3$.
\begin{figure*}[t]
\centering
\includegraphics[width=0.48\textwidth]{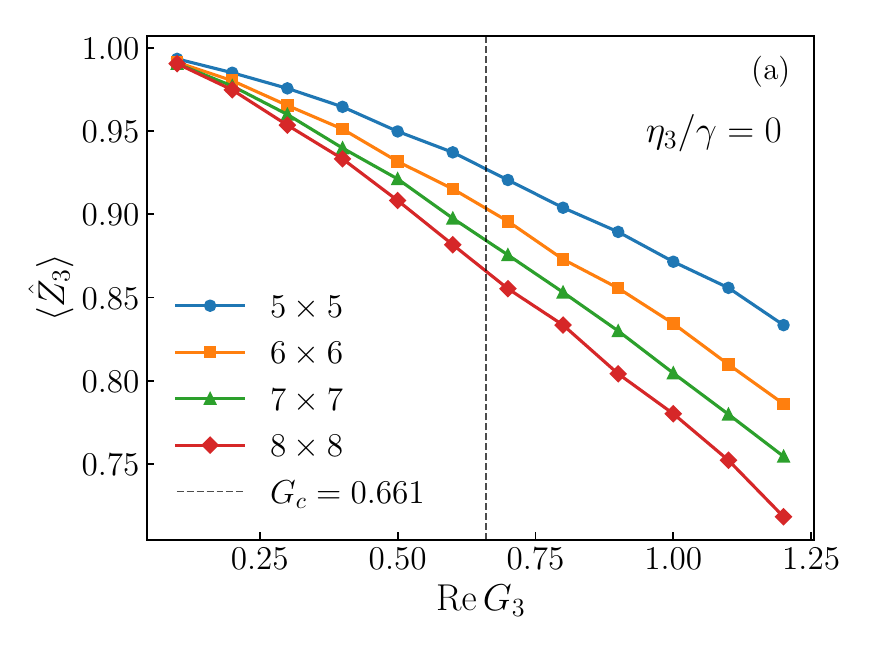}\hfill
\includegraphics[width=0.48\textwidth]{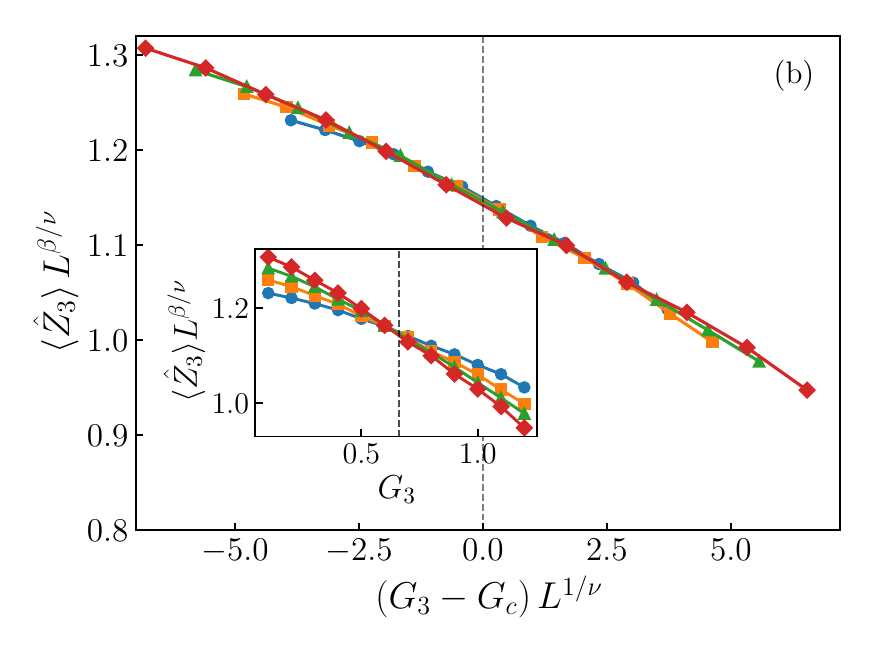}
\caption{
(a) Steady-state $\langle \hat{Z}_3 \rangle$ as a function of the real part of the three-photon drive $\mathrm{Re}\,G_3$ for 2D lattice sizes $5\!\times\!5$ to $8\!\times\!8$, computed with a total of $N_G = 24$ Gaussian components. Here $\langle \hat{Z}_3 \rangle$ is the $\mathbb{Z}_3$ parity order parameter, which equals unity in the symmetric phase and vanishes as the three symmetry-broken sectors become equally populated. The increasingly sharp drop with system size signals a continuous $\mathbb{Z}_3$ symmetry-breaking transition. (b) Finite-size scaling collapse using the critical exponents of the 2D classical 3-state Potts model, $\beta=1/9$ and $\nu=5/6$, with fitted critical drive $G_c = 0.661$, demonstrating that the nonequilibrium transition belongs to the classical three-state Potts universality class. Inset: rescaled parity $\langle \hat{Z}_3 \rangle\,L^{\beta/\nu}$ as a function of the unrescaled three-photon drive $G_3$, showing the size-independent crossing at $G_c$. System parameters: $J/\gamma = 20$, $U/\gamma =40$, $\Delta/\gamma = -20$, $\eta_3/\gamma =0$.
}
\label{fig:Z3_2d}
\end{figure*}
The Lindbladian $\mathcal{L}$ admits a 
$\mathbb{Z}_3$ symmetry and is invariant under 
$\hat{a}_j \to e^{i2\pi k/3}\hat{a}_j,~k\in \mathbb{Z}$. As the drive $G_3$ is increased, this 
symmetry is spontaneously broken: the system transitions from the 
vacuum to an ordered steady state. In the strong-driving 
limit $|G_3| \gg \gamma, \eta_3$, this steady state is spanned by three 
coherent states 
$|\alpha^{(k)}\rangle = |\alpha e^{i2\pi k/3}\rangle$, 
$k = 0, 1, 2$~\cite{Minganti2023}, and the exact steady state of a 
finite system, respecting the $\mathbb{Z}_3$ symmetry, is the equal-weight mixture
\begin{equation}\label{eq:rho_ss}
\hat{\rho}_{\rm ss} = \dfrac{1}{3}\sum_{k=0}^{2}
|\alpha^{(k)}\rangle\langle\alpha^{(k)}|\,.
\end{equation}
The system undergoes thus a spontaneous symmetry breaking of the $\mathbb Z_3$ symmetry. 
A natural order parameter is the $\mathbb{Z}_3$ parity 
operator, defined as 
\begin{equation}\label{eq:Z3_parity}
\hat{Z}_3 = \exp\!\biggl(\frac{i2\pi}{3}
\sum_j \hat{a}_j^\dagger \hat{a}_j\biggr).
\end{equation}
In the symmetric 
(vacuum) phase $\langle \hat{Z}_3 \rangle = 1$, while in the 
broken-symmetry phase all three $\mathbb{Z}_3$ sectors become 
equally populated and $\langle \hat{Z}_3 \rangle \to 0$ in the 
thermodynamic limit. This parallels the behavior of the 
$\mathbb{Z}_2$ parity 
$\hat{\Pi} = e^{i\pi \sum_j  \hat{a}^\dagger_j \hat a_j}$ in the two-photon driven-dissipative Bose-Hubbard
model~\cite{Rota2019, Verstraelen2020, Tosca2025VMG}.

In the ordered phase, a mean-field picture predicts that each site is occupied by one of the three 
coherent states $|\alpha^{(k_j)}\rangle$, $k_j \in \{0,1,2\}$, 
where the amplitude $\alpha$ is determined by the 
mean-field steady-state equation. The effective energy of the 
lattice can be obtained by evaluating the expectation value of 
the Hamiltonian~\eqref{eq:Hamiltonian} on the product state 
$\bigotimes_j |\alpha^{(k_j)}\rangle$. Since all on-site terms 
are invariant under $k_j \to k_j + 1$, only the hopping 
contribution depends on the configuration $\{k_j\}$:
\begin{equation}\label{eq:Heff}
E_{\rm eff}(\{k_j\}) = -\frac{2J|\alpha|^2}{z}
\!\sum_{\langle j,j'\rangle}
\cos\!\left[\tfrac{2\pi}{3}(k_j - k_{j'})\right].
\end{equation}
Equation~\eqref{eq:Heff} has the structure of the $q$-state 
Potts model, the canonical generalization of the Ising model in 
classical statistical mechanics~\cite{Wu1982}. It describes 
classical spins $s_j \in \{1,\dots,q\}$ on a lattice, governed 
by the ferromagnetic Hamiltonian
\begin{equation}\label{eq:Potts}
H_{\rm Potts} = -\tilde{J}\!\sum_{\langle j,j'\rangle}
\delta_{s_j,s_{j'}}, \qquad \tilde{J}>0,
\end{equation}
which lowers the energy whenever neighboring spins take the same 
value and reduces to the Ising model for $q=2$. Identifying the 
spin $s_j$ with the $\mathbb{Z}_3$ index $k_j$, the effective 
energy~\eqref{eq:Heff} (see the derivation in the End Matter) 
is exactly the $q=3$ case with coupling 
$\tilde{J} = 2J|\alpha|^2/z > 0$, its cosine form being the 
equivalent three-state \emph{clock} representation of 
Eq.~\eqref{eq:Potts}. While Eq.~\eqref{eq:Heff} is derived only in the strong-driving limit, it suggests that the nonequilibrium phase transition should belong to the universality class of the 3-state Potts model. This expectation extends well beyond the validity of the effective Hamiltonian, since universality is dictated by the exact $\mathbb{Z}_3$ symmetry of the Liouvillian rather than by microscopic details. In 2D, the corresponding critical exponents of the classical $3$-state Potts model are $\beta = 1/9$ and $\nu = 5/6$~\cite{Wu1982,Cardy1996}.

\textit{Numerical method.}\,---\,We simulate the cubically-driven Bose-Hubbard lattice using the 
Variational Multi-Gaussian (VMG) method~\cite{Tosca2025VMG}, which 
represents the many-body Wigner function as a sum of $N_G$ Gaussian 
components in phase space, evolved via the Dirac--Frenkel variational 
principle, using automatic differentiation to compute the equations of motion.
Each variational component is represented by the real part of a multimode Gaussian $G(\boldsymbol{\xi};\boldsymbol{\theta}_i)$ in phase space $\boldsymbol{\xi}\in\mathbb{R}^{2M}$, specified by a center $\boldsymbol{\mu}\in\mathbb{C}^{2M}$, a block-diagonal covariance whose $j$-th $2\times2$ block is a single-mode squeezed vacuum with magnitude $r_j$ and angle $\theta_j$, and a real (possibly negative) weight $c$; inter-site correlations are thus built up by the superposition of components rather than encoded within the covariances of individual Gaussians. The $N_G$ Gaussians are organized into $N_G/3$ $\mathbb{Z}_3$-symmetric triples, so that the full variational ansatz for the many-body Wigner function reads
\begin{equation}\label{eq:VMG_Z3_ansatz}
W_{\boldsymbol{\theta}}(\boldsymbol{\xi}) \;=\; \frac{1}{3}\sum_{i=1}^{N_G/3}\sum_{k=0}^{2} \Re[G\!\left(\boldsymbol{\xi};\,\mathcal{R}^k\boldsymbol{\theta}_i\right)],
\end{equation}
where $\mathcal{R}$ denotes the action of the $2\pi/3$ phase-space rotation on the parameters of a Gaussian: the center transforms as $\boldsymbol{\mu}\to R_{2\pi/3}\,\boldsymbol{\mu}$ and the squeezing angles shift as $\theta_j\to\theta_j+4\pi/3$. 
%Note that the covariance ellipses rotating at twice the phase-space rate, while $r_j$ and $c$ are left invariant.
The order parameter $\langle\hat{Z}_3\rangle$ is obtained from its Weyl symbol [Eq.~\eqref{eq:Zn_Weyl_multi}, End Matter], which reduces it to a sum of analytical Gaussian integrals over the ansatz, manifestly real by construction.

We benchmark the $\mathbb{Z}_3$-symmetric VMG ansatz against exact numerical simulations of the single-mode cubically driven Kerr parametric oscillator. Figure~\ref{fig:VMG_benchmark}(a) shows a snapshot in time of the VMG Wigner function at $\eta_3 t = 10$ with $N_G = 36$ (twelve Gaussian triples), in a regime of strong three-photon driving and appreciable three-photon losses. The reconstructed state exhibits the expected three-fold phase-space structure together with pronounced Wigner negativities, highlighting its strongly non-Gaussian and genuinely quantum character. The pointwise deviation from the exact Wigner function, shown in Fig.~\ref{fig:VMG_benchmark}(b), remains roughly two orders of magnitude smaller than the Wigner function itself, demonstrating that the VMG ansatz accurately captures the dissipative dynamics even in this strongly non-Gaussian regime, where standard stochastic phase-space approaches become unfeasible.
\begin{figure*}[t]
\centering
\includegraphics[width=0.48\textwidth]{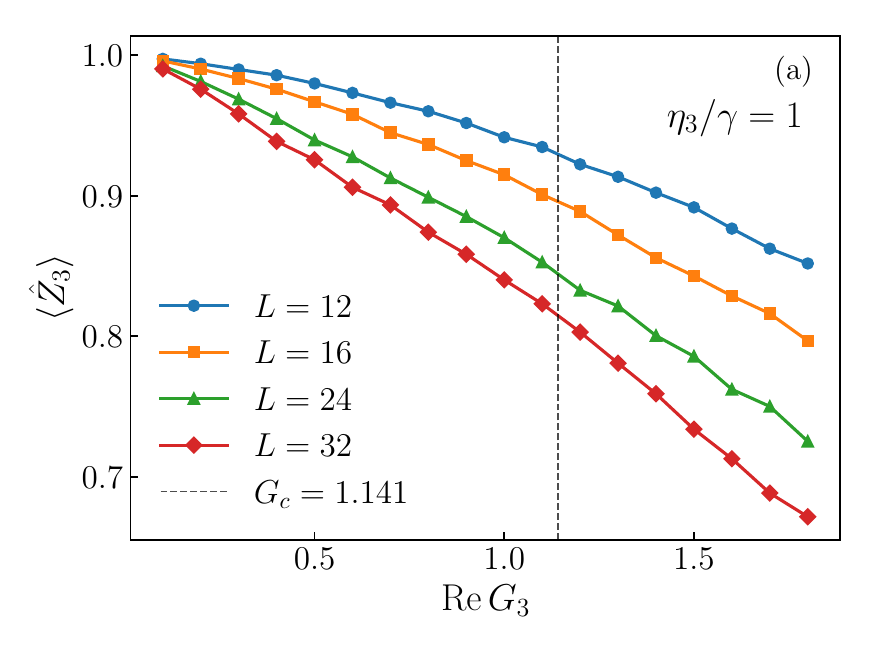}\hfill
\includegraphics[width=0.48\textwidth]{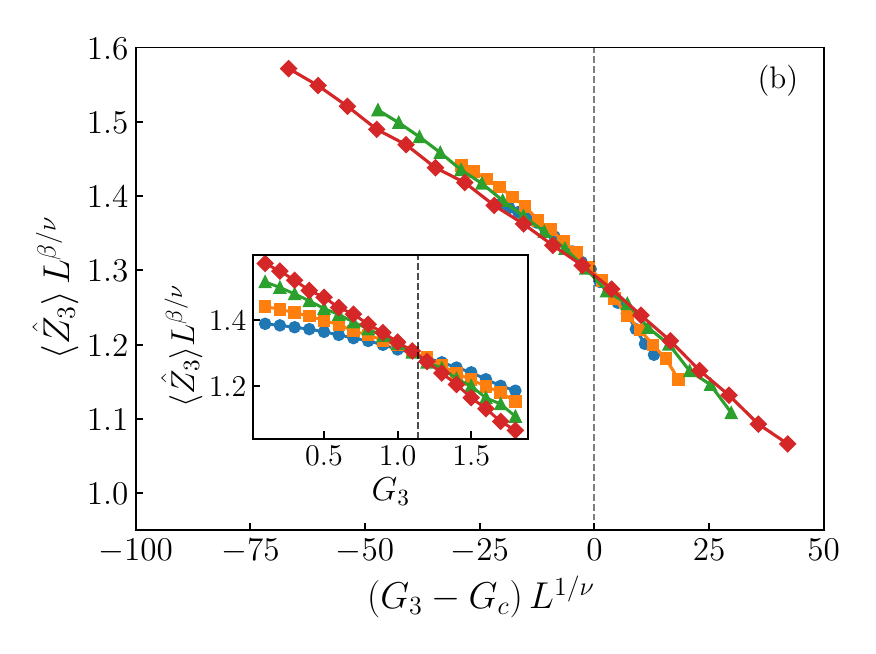}
\caption{
(a) Steady-state of the order parameter $\langle \hat{Z}_3 \rangle$ as a function of the real part of the three-photon drive $\mathrm{Re}\,G_3$ for 1D chains of length $L = 12, 16, 24, 32$, computed with $N_G = 30$ Gaussian components. The steady-state values are obtained by averaging the late-time dynamics over a time window of $5\,\gamma^{-1}$ ($500$ time steps). (b) Finite-size scaling collapse using the critical exponents of the $1$D quantum 3-state Potts universality class---equivalently, those of the 2D classical 3-state Potts model, $\beta=1/9$ and $\nu=5/6$---with fitted critical drive $G_c=1.141$, demonstrating that the nonequilibrium transition is governed by quantum Potts criticality. Inset: rescaled parity $\langle \hat{Z}_3 \rangle\,L^{\beta/\nu}$ as a function of the unrescaled three-photon drive $G_3$, showing the size-independent crossing at $G_c$. System parameters: $J/\gamma = 20$, $U/\gamma =40$, $\Delta/\gamma = -20$, $\eta_3/\gamma =1$.
}
\label{fig:Z3_1d}
\end{figure*}
\begin{table}[t!]
\centering
\caption{Universality classes of symmetry-breaking transitions in $n$-photon driven-dissipative bosonic lattices. In all cases the system is subject to single-photon losses, while the presence or absence of $n$-photon losses determines the nature of the criticality. The table illustrates the emerging correspondence whereby $n$-photon driving fixes the discrete $\mathbb{Z}_n$ symmetry and the associated universality class, whereas $n$-photon losses promote the transition from classical to quantum criticality.\label{tab:universality}}
\begin{ruledtabular}
\begin{tabular}{c c c l l}
Symm. & $n$-loss & $d$ & Universality class & Ref. \\
\hline
$\mathbb{Z}_2$ & absent  & 2 & 2D classical Ising      & \cite{Verstraelen2020}\\
$\mathbb{Z}_2$ & present & 1 & $1$D quantum Ising  & \cite{Rota2019}\\
$\mathbb{Z}_2$ & present & 2 & $2$D quantum Ising  & \cite{Rota2019}\\
$\mathbb{Z}_3$ & absent & 2 & $2$D classical Potts  & this work\\
$\mathbb{Z}_3$ & present & 1 & $1$D quantum Potts  & this work\\
\end{tabular}
\end{ruledtabular}
\end{table}

\textit{Universality class.}\,---\,We now present numerical evidence that the $\mathbb{Z}_3$ symmetry-breaking transition is continuous and falls into the 3-state Potts universality class, and we examine the qualitative role played by three-photon losses in setting the nature of the criticality. Near the critical point, the order parameter satisfies the standard finite-size scaling form~\cite{Fisher1972, Cardy1996}
\begin{equation}\label{eq:scaling}
\langle \hat{Z}_3 \rangle = L^{-\beta/\nu}\, 
\mathcal{F}\!\bigl[(G_3 - G_c)\,L^{1/\nu}\bigr],
\end{equation}
where $\mathcal{F}$ is a universal scaling function and $L$ the linear lattice size. We perform two complementary finite-size studies, in 2D and 1D, that together establish the full picture.

We first consider the model on 2D square lattices with periodic boundary conditions and only single-photon losses ($\eta_3 = 0$). Figure~\ref{fig:Z3_2d}(a) shows the steady-state $\langle \hat{Z}_3 \rangle$ as a function of $\mathrm{Re}\,G_3$ for lattice sizes from $5\times 5$ to $8\times 8$. At weak driving the system sits in the symmetric phase with $\langle \hat{Z}_3 \rangle \approx 1$; increasing $G_3$ progressively populates the three-fold degenerate manifold and drives $\langle \hat{Z}_3 \rangle$ towards lower values. The sharpening of the drop with increasing $L$ signals the approach to a phase transition in the thermodynamic limit. Figure~\ref{fig:Z3_2d}(b) shows the scaling collapse $\langle \hat{Z}_3 \rangle\, L^{\beta/\nu}$ versus $(G_3-G_c)\,L^{1/\nu}$ obtained with the exact 2D classical 3-state Potts exponents $\beta=1/9 $, $\nu=5/6 $~\cite{Wu1982} and the fitted critical pump $G_c=0.661$. The four lattice sizes collapse onto a single universal curve, confirming that --- in the absence of three-photon losses --- the symmetry-breaking transition belongs to the 2D classical 3-state Potts universality class. Crucially, this shows that a driven-dissipative bosonic lattice with three-photon parametric driving can display the criticality of the $\mathbb{Z}_3$ Potts ferromagnet.

We next switch on three-photon losses ($\eta_3/\gamma=1$) and study a 1D chain. Figure~\ref{fig:Z3_1d} shows the analogous data for chain lengths $L=12,16,24,32$. As in the 2D case, $\langle \hat{Z}_3 \rangle$ drops continuously with $G_3$ and the drop sharpens with $L$, with critical pump $G_c=1.141$. The scaling collapse [Fig.~\ref{fig:Z3_1d}(b)] is now performed with the critical exponents of the 1D quantum 3-state Potts universality class which, by the standard quantum-to-classical correspondence, coincide with those of the 2D classical Potts model, $\beta=1/9$ and $\nu=5/6$~\cite{Wu1982,Cardy1996}. The collapse demonstrates that the 1D chain in the presence of three-photon losses is in the quantum critical regime, promoting the criticality from classical to  quantum \footnote{The largest system size considered is $L=32$, compared with $L=64$ in Ref.~\cite{Rota2019}, because automatic differentiation requires storing sixth-order derivatives generated by the three-photon-loss Lindbladian, leading to substantially higher memory requirements. Despite the smaller system sizes, the quality of the finite-size scaling collapse remains comparable to that reported in Ref.~\cite{Rota2019}.}.

The results obtained in 1D and 2D reveal a broader organizing principle for driven dissipative bosonic lattices with multiphoton processes. For $\mathbb{Z}_2$ symmetry, the quadratically driven Bose-Hubbard model realizes a classical Ising critical point in the absence of two-photon losses~\cite{Verstraelen2020}, whereas the inclusion of two-photon losses drives the system into the quantum Ising universality class in the corresponding dimension of the system~\cite{Rota2019}. Figures~\ref{fig:Z3_2d} and~\ref{fig:Z3_1d} demonstrate that the same mechanism extends to $\mathbb{Z}_3$ symmetry. Without three photon losses, the transition belongs to the classical three state Potts universality class, whereas three photon losses promote it to the corresponding quantum universality class. Table~\ref{tab:universality} summarizes these results. Together, the $\mathbb{Z}_2$ and $\mathbb{Z}_3$ cases suggest a general correspondence: the order of the multiphoton drive determines the underlying discrete symmetry and its associated universality class, while the corresponding multiphoton losses determine whether the emergent critical behavior is classical or quantum.

\textit{Conclusion.}\,---\,We have demonstrated that driven-dissipative bosonic lattices with a broken $\mathbb{Z}_3$ symmetry can display emergent three-state Potts criticality, encoded in the steady state of the $\mathbb{Z}_3$ photon-number parity operator. Using finite-size scaling analyses based on the Variational Multi-Gaussian (VMG) approach, we have established that the cubically driven-dissipative Bose-Hubbard model realizes the classical 2D three-state Potts universality class on 2D lattices with only single-photon losses, and the 1D quantum three-state Potts universality class on 1D  chains where three-photon losses are also present.

These results provide the first identification of Potts criticality beyond the extensively studied $\mathbb{Z}_2$ case in driven-dissipative bosonic lattices. More broadly, they reveal how equilibrium universality classes can emerge from genuinely nonequilibrium many-body dynamics, extending the connection between open quantum systems and equilibrium critical phenomena from the Ising to the Potts paradigm.

Taken together with previous studies of quadratically driven $\mathbb{Z}_2$ lattices, our results suggest a general correspondence in multiphoton driven-dissipative bosonic systems: the order of the multiphoton drive determines the underlying $\mathbb{Z}_n$ symmetry and its associated universality class, while the corresponding multiphoton losses promote the transition from classical to quantum criticality. An intriguing open question is whether this correspondence extends to higher discrete symmetries. In particular, four-photon driven lattices are natural candidates to realize the Ashkin--Teller universality class~\cite{Ashkin1943}, opening the way to the exploration of a broader hierarchy of equilibrium universality classes in intrinsically nonequilibrium quantum systems.

\begin{acknowledgments}
We acknowledge support from the grant ANR-24-RRII-0001 Polaritonic from the French Government managed by the ANR under the France 2030 programme. This work was granted access to the HPC resources of IDRIS under the allocation A0190916893 made by GENCI. We acknowledge Luca Giacomelli and Francesco Carnazza for discussions and their contributions to the development of the VMG method. 
\end{acknowledgments}

\bibliography{biblio}

\onecolumngrid
\vspace{1.5em}
\begin{center}
\textbf{\large End Matter}
\end{center}
\vspace{0.5em}
\twocolumngrid

\setcounter{equation}{0}
\renewcommand{\theequation}{A\arabic{equation}}

\textit{Effective Potts energy in the ordered phase.}\,---\,We derive the effective lattice energy of Eq.~\eqref{eq:Heff}. In the strong driving ordered phase, each site occupies one of the three coherent states
$|\alpha^{(k_j)}\rangle = |\alpha\,e^{i2\pi k_j/3}\rangle$,
with $k_j \in \{0,1,2\}$, and the many body state is the product
$|\Psi\rangle = \bigotimes_j |\alpha^{(k_j)}\rangle$. Defining
$\phi = 2\pi/3$, we evaluate
$E_{\rm eff}(\{k_j\}) = \langle\Psi|\hat{H}|\Psi\rangle$
using $\hat{a}_j|\Psi\rangle = \alpha\,e^{i\phi k_j}|\Psi\rangle$.

The onsite terms are independent of the configuration $\{k_j\}$. The number and Kerr terms,
\begin{equation}\label{eq:onsite_kerr}
\langle \hat{a}_j^\dagger \hat{a}_j\rangle = |\alpha|^2,
\qquad
\langle \hat{a}_j^{\dagger 2}\hat{a}_j^2\rangle = |\alpha|^4,
\end{equation}
are phase independent, while the cubic drive acquires the phase factor
$e^{\pm i3\phi k_j}$. Since $3\phi = 2\pi$ and $k_j\in\mathbb{Z}$,
\begin{equation}\label{eq:onsite_drive}
\langle G_3\hat{a}_j^{\dagger 3} + G_3^*\hat{a}_j^3\rangle
= G_3(\alpha^*)^3 e^{-i2\pi k_j} + \text{c.c.}
= 2\,\mathrm{Re}\!\left[G_3(\alpha^*)^3\right],
\end{equation}
which is again independent of $k_j$. Therefore, every onsite contribution reduces to a constant offset $E_0$, which determines the coherent state amplitude $\alpha$ through the single-site meanfield equation but does not distinguish between different configurations.

The only contribution that depends on the configuration arises from the hopping term. Using
$\langle\hat{a}_j^\dagger\hat{a}_{j'}\rangle
= |\alpha|^2 e^{i\phi(k_{j'}-k_j)}$, we find 
\begin{equation}\label{eq:hopping_exp}
\langle\hat{a}_j^\dagger\hat{a}_{j'}
+ \hat{a}_{j'}^\dagger\hat{a}_j\rangle
= 2|\alpha|^2\cos\!\left[\tfrac{2\pi}{3}(k_j - k_{j'})\right].
\end{equation}
Substituting Eq.~\eqref{eq:hopping_exp} into the hopping term of Eq.~\eqref{eq:Hamiltonian} and discarding the constant offset $E_0$ gives Eq.~\eqref{eq:Heff},
\begin{equation}\label{eq:Heff_app}
E_{\rm eff}(\{k_j\}) = -\frac{2J|\alpha|^2}{z}
\!\sum_{\langle j,j'\rangle}
\cos\!\left[\tfrac{2\pi}{3}(k_j - k_{j'})\right],
\end{equation}
which is equivalent to the ferromagnetic three state Potts Hamiltonian with coupling $\tilde{J} = 2J|\alpha|^2/z > 0$, and therefore belongs to the three state Potts universality class.

\setcounter{equation}{0}
\renewcommand{\theequation}{B\arabic{equation}}

\textit{Weyl symbol of the $\mathbb{Z}_n$ parity operator.}\,---\,We derive the Weyl symbol of $\hat{Z}_n = e^{i\phi\hat{n}}$ 
($\phi = 2\pi/n$, $\hat{n} = \hat{a}^\dagger\hat{a}$) for a 
single mode; the $M$-mode result factorizes. The rotation 
operator $\hat{R}(\phi) = e^{i\phi\hat{n}}$ has the position 
representation~\cite{Walls1994}
\begin{equation}\label{eq:Mehler}
\langle q' | \hat{R}(\phi) | q \rangle 
= \frac{e^{i\phi/2}}{\sqrt{2\pi i \sin\phi}}\,
e^{\frac{i}{2\sin\phi}[(q'^2+q^2)\cos\phi-2q'q]},
\end{equation}
valid for $\phi \neq k\pi$ ($n \geq 3$). The $n\!=\!2$ case 
recovers $\delta(q)\delta(p)$~\cite{Tosca2025VMG} as a 
distributional limit.

The Weyl symbol reads
\begin{equation}\label{eq:Weyl_def}
(Z_n)_W(q,p) = \!\int\! dy\,
\langle q\!-\!\tfrac{y}{2} | \hat{R}(\phi) 
| q\!+\!\tfrac{y}{2} \rangle\, e^{ipy}.
\end{equation}
Substituting~\eqref{eq:Mehler}, the exponent in the Mehler 
kernel simplifies to 
$-iq^2\tan(\phi/2) + iy^2/[4\tan(\phi/2)]$, 
after using standard half-angle identities. The remaining 
Gaussian integral over $y$ yields
\begin{equation}\label{eq:Zn_Weyl}
(Z_n)_W(q,p) = \frac{e^{i\phi/2}}{\cos(\phi/2)}\,
e^{-i(q^2 + p^2)\tan(\phi/2)}\,,
\end{equation}
an oscillatory function in phase space with frequency 
$\tan(\phi/2)$.

For $M$ modes, the Weyl symbol factorizes:
\begin{equation}\label{eq:Zn_Weyl_multi}
(Z_n^{\rm tot})_W = 
\biggl(\frac{e^{i\phi/2}}{\cos(\phi/2)}\biggr)^{\!M}
e^{-i\tan(\phi/2)\sum_j(q_j^2 + p_j^2)}.
\end{equation}
Within the VMG framework, $\langle \hat{Z}_n \rangle$ reduces 
to a sum of analytical Gaussian 
integrals~\cite{Tosca2025VMG}.

For the case $n = 3$ relevant to this work 
($\phi = 2\pi/3$), Eq.~\eqref{eq:Zn_Weyl} gives
\begin{equation}\label{eq:Z3_Weyl}
(Z_3)_W(q,p) = 2\,e^{i\pi/3}\,
e^{-i\sqrt{3}\,(q^2 + p^2)}\,,
\end{equation}
using $\cos(\pi/3) = 1/2$ and $\tan(\pi/3) = \sqrt{3}$.

\end{document}

%% file: shortcuts.tex
\let\originalleft\left
\let\originalright\right
\renewcommand{\left}{\mathopen{}\mathclose\bgroup\originalleft}
\renewcommand{\right}{\aftergroup\egroup\originalright}

\makeatletter
\newcommand*\bigcdot{{\color{gray}\mathpalette\bigcdot@{1.}}}
\newcommand*\bigcdot@[2]{\mathbin{\vcenter{\hbox{\scalebox{#2}{$\m@th#1\bullet$}}}}}
\makeatother

\newcommand{\bea}{\begin{equation}\begin{aligned}}
		\newcommand{\eea}{\end{aligned}\end{equation}}
\newcommand{\be}{\begin{equation}}
	\newcommand{\ee}{\end{equation}}